\documentclass[aps,floatfix]{revtex4}

\usepackage{amsmath}
\usepackage{graphicx}
\usepackage{epsfig}
\usepackage{amsfonts}
\usepackage{amssymb}

\begin{document}

\title{Two-bands superconductivity with intra- and interband pairing for synthetic superlattices}

\author{Alex A.\ Schmidt$^1$}
\author{Jose J.\ Rodr\'{\i}guez-N\'u\~nez$^2$}
\author{Antonio Bianconi$^3$}
\author{Andrea Perali$^4$}

\affiliation{$^1$Departamento de Matem\'atica, UFSM, 97105-900 Santa Maria, RS, Brazil \\ 
$^2$Lab.\ SUPERCOMP, Departamento de F\'{\i}sica--FACYT, University of Carabobo, Valencia, 
Venezuela \\ $^3$Physics Department, Sapienza Universit$\grave{a}$ di Roma, I-00185 Roma, Italy \\
$^4$School of Pharmacy, Physics Units, Universit$\grave{a}$ di Camerino, 
Via Madonna delle Carceri, I-62032, Camerino (MC), Italy}

\begin{abstract}
We consider a model for superconductivity in a two-band superconductor, having an anisotropic 
electronic structure made of two partially overlapping bands with a first hole-like and a second 
electron-like fermi surface. In this pairing scenario, driven by the interplay between 
interband $V_{i,j}$ and intraband $V_{i,i}$ pairing terms, we have solved the two gap equations 
at the critical temperature $T = T_c$ and calculate $T_c$ and the chemical potential $\mu$ as a 
function of the number of carriers $n$ for various values of pairing interactions, $V_{1,1}$, 
$V_{2,2}$, and $V_{1,2}$. The results show the complexity of the physics of condensates with 
multiple order parameters with the chemical potential near band edges.
\end{abstract}

\maketitle

\section{Introduction}\label{s1}

In a metallic system where both a first band  $V_{1}$ and a second band $V_{2}$, with no 
hybridization, cross the Fermi level the pairing is described by a matrix where the diagonal 
elements are the intraband Cooper pairing processes, namely, $V_{1,1}$ and $V_{2,2}$ and the 
non-diagonal elements, namely, $V_{1,2}$ that describes the interband pairing process 
associated with the transfer of pairs between the bands \cite{SMW,Kondo,Leggett}.

Today the open problem in condensed matter physics is the interband pairing where the 
standard BCS approximation that considers the Fermi level far away form the energy of 
band edges or critical points breaks down. In fact where the chemical potential is tuned 
near an electronic topological transition the high temperature superconductivity has been 
associated with quantum interference effects, called shape resonances, in the interband 
pairing terms since 1993 \cite{bianconi94,bianconi05}.

The shape resonance in the superconducting gaps is realized by material design of superconducting 
hetero-structures at atomic limit made of superlattices of superconducting layers with a relevant 
transversal hopping between layers \cite{bianconi93,pnictides1}. The material architecture and 
the superlattice misfit strain \cite{agrestini,poccian,pnictides2} are key terms for fine tuning 
the system to quantum critical points\cite{agrestini2,agrestini3} driven by electronic topological 
transitions.

At the time of discovery of high Tc cuprates in 1986, these materials have been considered to be 
isotropic 3D perovskites but later it was recognized that all high temperature superconducting 
cuprates are made of a superlattice of atomic copper oxide layers intercalated by different spacers 
where the evolution of the complex electronic structure with chemical doping has remained mysterious 
for many years. The large majority of scientific community has interpreted the available unconventional 
experimental data in terms of an effective single band for high temperature superconducting models. 
Few authors have interpreted the data in terms of mainly two non-hybridized bands and two Fermi 
surfaces at the nodal and anti-nodal point of the band structure \cite{bianconi93,kristoffel}. 
This key point for the physics of cuprates has been recently accepted by the large scientific 
community in 2010 driven by compelling experimental data \cite{norman10}.

Later different practical realizations of superlattices of superconducting atomic layers 
intercalated by spacers tuned at a shape resonance \cite{bianconi93} have been found in 2001 
in diborides \cite{bianconi01,bianconi02,Bussmann03,Ummarino04,Tsuda05,Cooley,Samuely,
carrington,Klie,sologubenko1,castro1,Putti_et_al,szabo1,giubileo1,pissas1,capua1} and in 
2008 in pnictides \cite{hosono,dagotto,chubukov}.

In these hetero-structures many experiments have provided experimental tests for multigap 
superconductivity in the clean limit also in the presence of many lattice impurities 
introduced to change the chemical potential and structural phase separation from nano-scale 
to micron-scale \cite{fratini,palmisano,caivano} with the signatures of complexity of 
condensed matter near a quantum critical point. The key point for high temperature 
superconductivity is the presence of shape resonances in the interband pairing due to 
pair transfer between condensates with different topological symmetry due to quantum 
interference effects between pairing channel. This scenario has been discussed using 
simple electronic models \cite{bianconi98,innocenti10} .

Superconducting pnictides \cite{hosono} are made of stacks of iron fcc atomic layers 
intercalated by spacers with a superlattice architecture similar to cuprates \cite{pnictides1} 
that show multigap superconductivity \cite{chubukov} with multiple bands crossing the Fermi 
level \cite{dagotto}.

While in diborides anomalous strong electron-phonon intraband interaction is quite 
relevant in two-band superconductivity scenario, in pnictides the intraband coupling 
is so small that has been neglected \cite{chubukov}. In fact experimental findings show 
that the scenario where interband pairing is the dominant interaction, proposed by some 
of us in previous works \cite{JJ-AAS,Rodriguez_2008}, has now an experimental test.

Here we investigate the case where the relative strength between the intraband and interband 
coupling is changing that is relevant for material design of novel high Tc superconductors 
made of synthetic superlattices. In these systems the relevant region for high temperature 
superconductivity is where the chemical potential is tuned between the band edge of the 
second band and the electronic topological transition  where the Fermi surface of the second 
band changes from  a 3D topology to a 2D topology with a corrugated cylindrical shape 
\cite{angilella,bianconi01}. Therefore we focus in a system with two similar overlapping 
bands such that the first hole-like (electron-like) Fermi surface coexists with a second 
electron-like (hole-like) Fermi surface.

The proposed model has the characteristic feature that for the occupation number of 1 electron 
per spin shows a critical point where the hole Fermi surface is the same as the electron Fermi 
surface (like in pnictides \cite{chubukov}) therefore tuning the chemical potential at this 
point there is a strong nesting between the two Fermi surfaces that determines a structural 
instability with a structural phase transition from tetragonal to orthorhombic phase \cite{pnictides2}. 
A scale invariant complex structure could appear in the proximity of this phase transition in 
pnictides \cite{pnictides2} as in cuprates \cite{fratini}.

We study a band structure that can be realized in superlattices of fcc lattice like in iron 
based multilayer materials that have provided a particular realization of multiband superconductivity 
in the clean limit with a high critical temperature.

In this paper we adopt the point of view that both the interband and intraband scattering terms 
are indeed very important to explain the response of the superconducting phase to the variation 
of the chemical potential. On the other side the case of pnictides is considered to be the case 
where intraband coupling can be neglected as we have proposed in the previous work before the 
discovery of pnictides \cite{Rodriguez_2008}. Now we focus our interest on the fact that it is 
important to consider different cases where both intra-band and interband pairing are relevant 
that will be relevant for novel synthetic superconducting superlattices. We consider the case where 
the interband pairing $V_{1,1}$ is as important as the intraband contributions, namely, $V_{2,2}, 
V_{1,2}$. Here we report the study of the variations of the critical temperature and gap ratio as 
a function of the charge density crossing electronic phase transitions. 

This paper is organized as follows. In Section \ref{s2} we present the four self-consistent equations: 
(\textbf{1}) two self-consistent equations (Eqs.\ (\ref{self-consistent1} - \ref{self-consistent2})) 
to be solved at $T = T_c$, namely, the equation for the superconducting critical temperature $T_c$ 
coupled to the equation of the chemical potential $\mu$ at a given particle density $n$, and 
(\textbf{2}) three self-consistent equations for the gap parameters at $T=0$ and the total density 
(Eqs.\ (\ref{number_carriers} - \ref{Delta_sigma})). In Section \ref{s4} we present our numerical results. 
Finally we conclude in Section \ref{s5}.

\section{The effective Hamiltonian and the $BCS$ equations}\label{s2}

\indent The model Hamiltonian we use is the following
\begin{eqnarray}\label{hamilton}
H - \mu N &=&
\sum_{\vec{k},\gamma}\varepsilon_{1}(\vec{k})\,
a^{\dagger}_{\vec{k},\gamma}a_{\vec{k},\gamma}
+\sum_{\vec{k},\gamma}\varepsilon_{2}(\vec{k})\,
b^{\dagger}_{\vec{k},\gamma}b_{\vec{k},\gamma} \nonumber \\
&-&V_{1,2}~\sum_{\vec{k},\vec{k'}}\left[a^{\dagger}_{\vec{k},\uparrow}
a^{\dagger}_{-\vec{k},\downarrow}
b_{\vec{k'},\uparrow}b_{-\vec{k'},\downarrow} + h.c. \right]\nonumber \\
&-&V_{1,1}~\sum_{\vec{k},\vec{k'}}~a^{\dagger}_{\vec{k},\uparrow}
a^{\dagger}_{-\vec{k},\downarrow}
a_{\vec{k'},\uparrow}a_{-\vec{k'},\downarrow}\nonumber \\ 
&-& 
V_{2,2}~\sum_{\vec{k},\vec{k'}}~b^{\dagger}_{\vec{k},\uparrow}
b^{\dagger}_{-\vec{k},\downarrow}
b_{\vec{k'},\uparrow}b_{-\vec{k'},\downarrow}
\end{eqnarray}
\noindent where $\gamma = \pm = \uparrow, \downarrow$ and
\begin{eqnarray}
\label{tightbindings}
\varepsilon_{1}(\vec{k}) &=& \epsilon_{1}(\vec{k})- \mu ~~~~~~~ \label{bandstructure1} \\
\varepsilon_{2}(\vec{k}) &=& \epsilon_{2}(\vec{k})- \mu ~~~~~~~ \label{bandstructure2}
\end{eqnarray}

\noindent are the three-dimensional tight-binding bands of bands 1 and 2. 
In previous papers \cite{Rodriguez_2008,JJ-AAS}, 
we have considered a Hamiltonian with $V_{1,2} = V$, and 
$V_{1,1} = V_{2,2} = 0$, as the interband 
and intraband pairing potentials, respectively. Here, these 
conditions are relaxed. We consider the simple case of 
a cubic lattice\cite{hexagonal}. The energy band dispersion is given by
\begin{eqnarray}\label{chosen_bandstructure}
\epsilon_{1}(\vec{k}) &=& \epsilon^0_{1} - 2\,t_{1}\,[\cos(k_x) + 
                        \cos(k_y) + s_1\cos(k_z)] ~~~~~~~ \label{bandstructure1a} \\
\epsilon_{2}(\vec{k}) &=& \epsilon^0_{2} - 2\,t_{2}\,[\cos(k_x) +
                        \cos(k_y) + s_2\cos(k_z)]~. ~~~~~~~ \label{bandstructure2a}
\end{eqnarray}

\noindent The Hamiltonian given in Eq.\ (\ref{hamilton}) is the same 
used in Ref.\ \cite{JJ-AAS}s.

Going through the same technical steps followed in Ref.\ \cite{JJ-AAS} 
we obtain the following self-consistent equations
\begin{eqnarray}
 1 - n &=& F_1(\mu,T_c) \label{self-consistent1} \\
1 &=& F_2(\mu,T_c,\omega_c) \label{self-consistent2} 
\end{eqnarray}
\noindent where
\begin{eqnarray}
F_1(\mu,T_c) &=& F_{1,1}(\mu,T_c) + F_{1,2}(\mu,T_c) \label{definitions1} \\[2mm]
F_2(\mu,T_c,\omega_c) &=& \left[ V^{2}_{1,2} - V_{1,1}\,V_{2,2} 
\right]\,F_{1}\,F_{2} + V_{1,1}\,F_{1} + V_{2,2}\,F_{2} \label{definitions2}
\end{eqnarray}
\noindent and
\begin{eqnarray}
F_{1,1}(\mu,T_c) &=& \int_0^{\pi}\,\dfrac{d\vec{k}}{2\,\pi^{3}}\,\,
\tanh\left[\frac{\varepsilon_1(\vec{k})}{2k_BT_c}\right]  \\[2mm]
F_{1,2}(\mu,T_c) &=& \int_0^{\pi}\,\dfrac{d\vec{k}}{2\,\pi^{3}}\,\,
\tanh\left[\frac{\varepsilon_2(\vec{k})}{2k_BT_c}\right]  \\[2mm]
F_{1} &=& \int_0^{\pi}\,\dfrac{d^{3}\vec{k}}{2\pi^{3}}\,\,\frac{\chi_{1}(\vec{k})}{\varepsilon_1(\vec{k})}\,\,
\tanh\left[\frac{\varepsilon_1(\vec{k})}{2k_BT_c}\right] \\[2mm]
F_{2} &=& \int_0^{\pi}\,\dfrac{d^{3}\vec{k}}{2\pi^{3}}\,\,\frac{\chi_{2}(\vec{k})}{\varepsilon_2(\vec{k})}\,\,
\tanh\left[\frac{\varepsilon_2(\vec{k})}{2k_BT_c}\right] 
\end{eqnarray}
\noindent where $k_B$ is the Boltzmann constant and $\chi_i(\vec{k}) = 1$ if 
$|\varepsilon_i(\vec{k})| \leq \omega_c$ and zero otherwise, $i =1 , 2$ and 
$\omega_c$ is an energy cutoff (the Debye frequency) for the effective interaction. 
We solve the $BCS$ equations at $T = 0\,K$. These
equations are:
\begin{eqnarray}
n &=& 1 - F_1^{'}(\mu) \label{number_carriers} \\
\Delta_1 &=& \Delta_1^*\,G_1 \label{Delta_sigma} \\
\Delta_2 &=& \Delta_2^*\,G_2 \label{Delta_pi}
\end{eqnarray}
\noindent where
\begin{eqnarray}
F_1^{'}(\mu) &=& \int_0^{\pi}\,\dfrac{d\vec{k}}{2\,\pi^{3}}\,\, 
\left[ \dfrac{\varepsilon_1(\vec{k})}{E_1(\vec{k})} + 
 \dfrac{\varepsilon_2(\vec{k})}{E_2(\vec{k})} \right] \label{def1} \\[2mm]
G_{i} &=& \int_0^{\pi}\,\dfrac{d\vec{k}}{2\pi^{3}}\,\,\frac{\chi_i(\vec{k})}{E_i(\vec{k})} 
~~~;~~~i = 1,2 ~~~\label{def2} \\[2mm]
E_i(\vec{k}) &=& \sqrt{\varepsilon_i(\vec{k})^2+(\Delta_i^*)^2} ~~~;~~~i = 1,2 ~~~\label{defekgap} \\[2mm]
\Delta_1^* &=&  V_{1,1}\,\Delta_1 + V_{1,2}\,\Delta_2 \nonumber \\[2mm]
\Delta_2^* &=&  V_{1,2}\,\Delta_1 + V_{2,2}\,\Delta_2 \nonumber 
\end{eqnarray}

Results for the numerical solution of the self-consistent Eqs.\ (\ref{self-consistent1}-\ref{Delta_sigma})
are presented in Section \ref{s4}.

\section{Searching for suitable pairing parameters and numerical results}\label{s4}

%\section{Searching for suitable pairing parameters}\label{s3}

\begin{figure}[htb] %  l   b   r   t
\includegraphics[trim=035 050 070 050,clip,
   width=0.75\textwidth,angle=0]{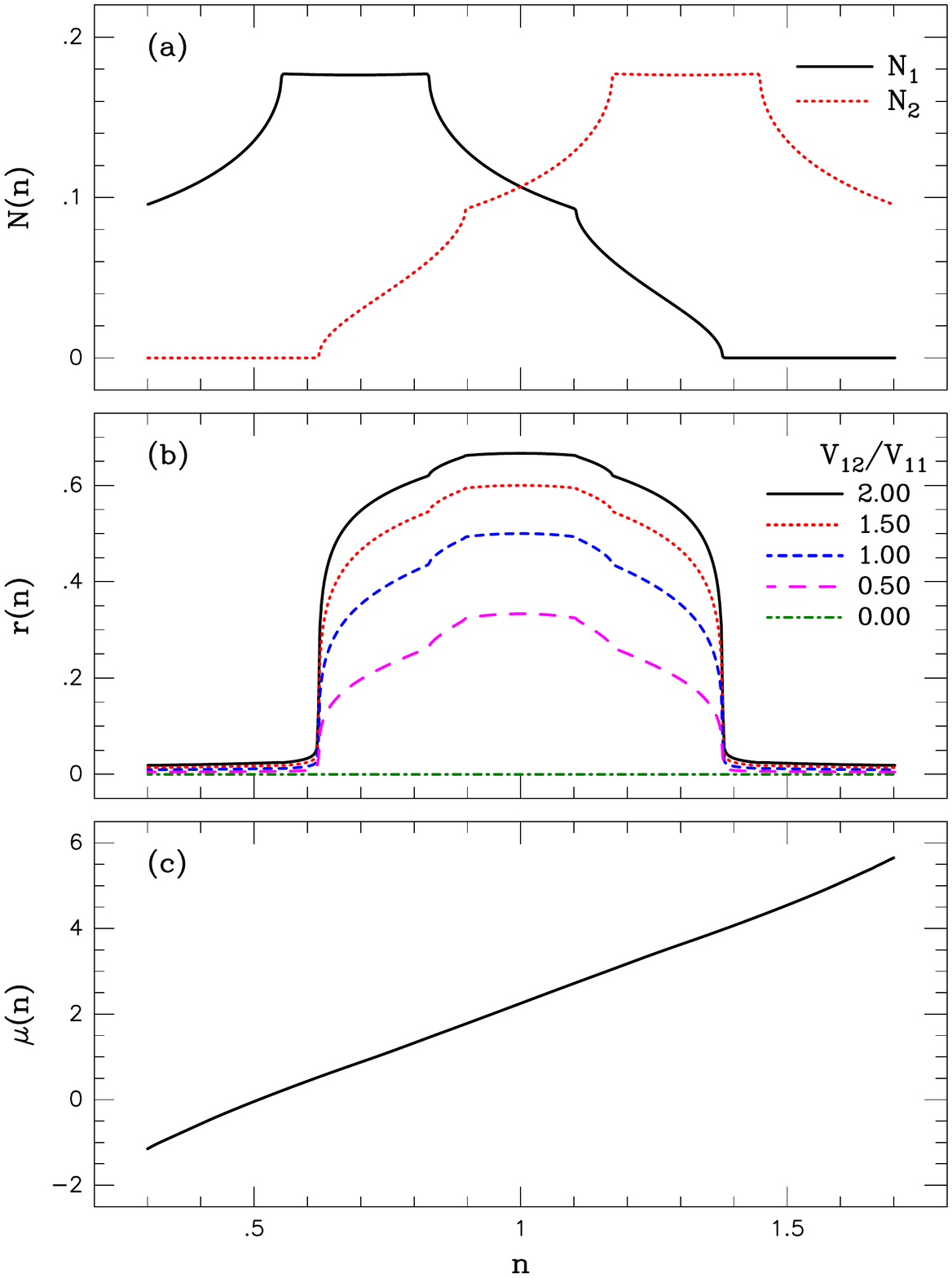}
\caption{Panel (a): the density of states $N_1$ and $N_2$ as a function of the carrier number $n$ 
according to the chosen band parameters $t_1 = t_2 = 1$ $eV$, $s_1 = s_2 = 0.5$, $\epsilon^0_1 = 0.0$ 
$eV$ and $\epsilon^0_2 = 4.5$ $eV$; panel (b): the ratio \(\displaystyle r(n) = V_{12}\,
\sqrt{N_1\,N_2}\,/\,\lambda_-\) for $V_{1,1} = V_{2,2}$  and $V_{1,2}/V_{1,1} = 0.0, 0.5, 
1.0, 1.5$ and $2.0$ (see Eq.\ (\ref{eq.lambda}) for the definition of $\lambda_-$), useful as a tracer 
of the region where the interband pairing is relevant; panel (c): the self-consistent numerical solution
of Eqs.\ (\ref{self-consistent1}-\ref{self-consistent2}) for chemical potential $\mu$ as a function of 
$n$ at $T = T_c$, which proved to be independent of the choice of $V_{1,1}, V_{2,2}, V_{1,2}$ and 
$\omega_c$ and matches to the $\mu\,\times\,n$  relation at $T = 0\,K$ obtained from Eqs.\ 
(\ref{number_carriers} -\ref{Delta_pi}) (coloured online).} \label{fig1}
\end{figure}

In Fig.\ \ref{fig1} the two density of states $N_1$ and $N_2$, corresponding to the energy dispersion 
of Eqs.\ (\ref{bandstructure1}-\ref{bandstructure2})  are presented in panel (a) as a function of 
the electron number density per spin $n$ for the band structure parameters chosen as
$t_1 = t_2 = 1$ $eV$, $s_1 = s_2 = 0.5$, $\epsilon^0_1 = 0.0$ $eV$ and $\epsilon^0_2 = 4.5$ $eV$,
{\it i.e.}, two similar overlapping bands so that the first hole-like (electron-like) Fermi surface 
coexists with a second electron-like (hole-like) Fermi surface. 

The region where the interband pairing is relevant is shown in panel (b) of Fig.\ \ref{fig1} 
(for $V_{1,1} = V_{2,2}$ and $V_{1,2}/V_{1,1} = 0.0, 0.5, 1.0, 1.5$ and $2.0$) by an effective 
density tracer given by the ratio \(\displaystyle r(n) = V_{12}\,\sqrt{N_1\,N_2}\,/\,\lambda_-\) 
where $\lambda_-$ is the effective density of states given by the standard $BCS$ theory as:
\begin{eqnarray}
T_c &=& 1.13~\omega_c \exp (\lambda_{\pm}) \label{critical} \nonumber
\end{eqnarray}
\noindent where
\begin{eqnarray}
\lambda_{\pm} &=& \dfrac
{2\,N_1\,N_2\,\left[V_{1,1}\,V_{2,2} - V^{2}_{1,2}\right]} 
{N_1\,V_{1,1} + N_2\,V_{2,2} \pm N} \label{eq.lambda} \\[2mm]
N &=& \sqrt {\left[N_1\,V_{1,1}-N_2\,V_{2,2}\right]^2 + 4\,N_1\,N_2\,V^2_{1,2}}~. \nonumber
\end{eqnarray}

Finally, in panel (c) of Fig.\ \ref{fig1} we present the self-consistent numerical solution 
of Eqs.\ (\ref{self-consistent1}-\ref{self-consistent2}) for chemical potential $\mu$ 
as a function of carrier density $n$ which presents a quasi-linear dependence with $n$.

\begin{figure}[htb] %  l   b   r   t
\includegraphics[trim=035 050 070 050,clip,
   width=0.75\textwidth,angle=0]{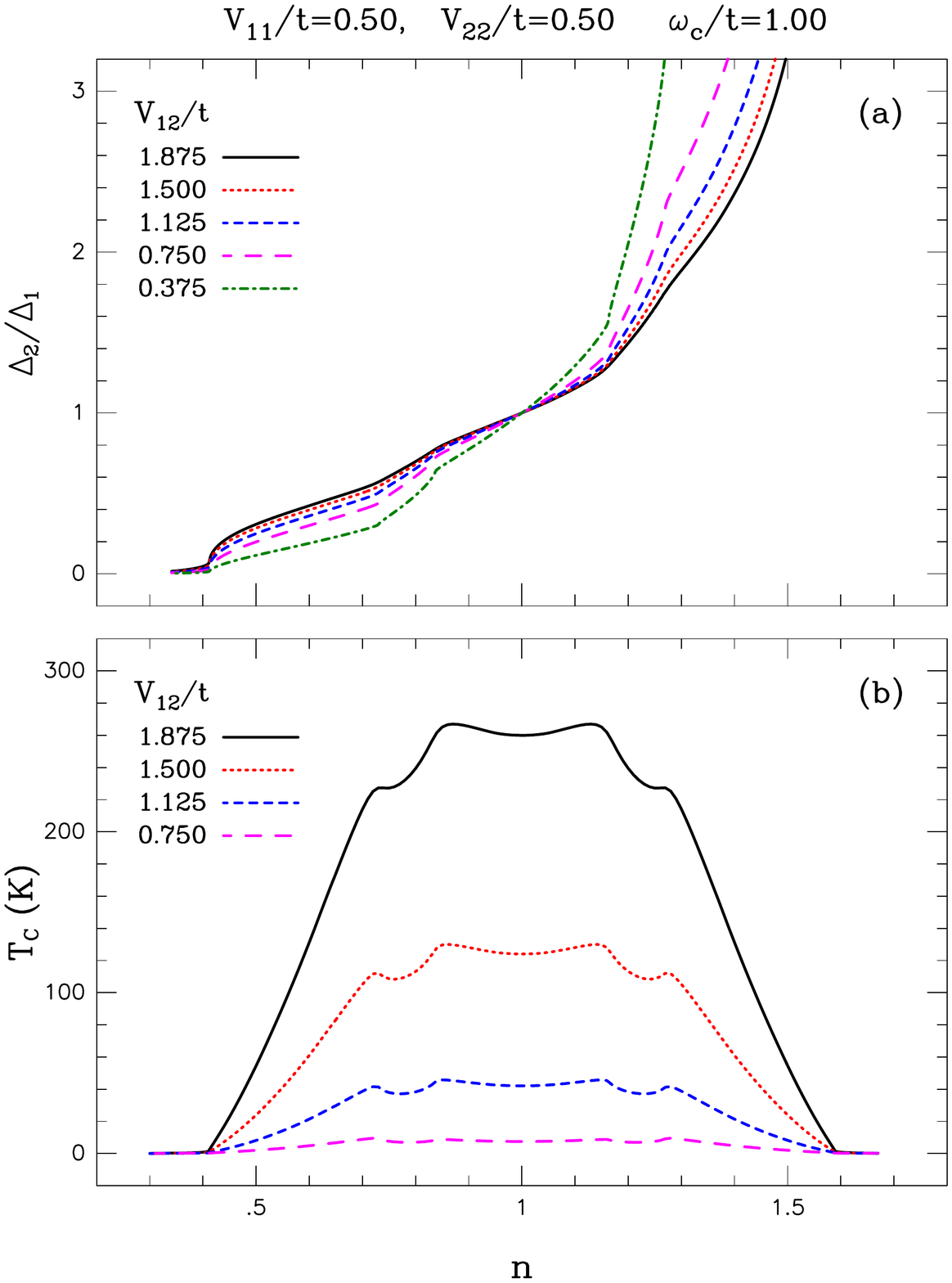}
\caption{For the given set of values $V_{1,1}/t = V_{2,2}/t = 0.50$, $\omega_c/t = 1.00$ and 
$V_{1,2}/t = 0.375, 0.750, 1.125, 1.500$ and $1.875$ the panel (a) shows the self-consistent 
numerical solution of Eqs.\ (\ref{number_carriers} -\ref{Delta_pi}) for the energy gap ratio 
$\Delta_2/\Delta_1$ at $T = 0\,K$ and panel (b) shows the self-consistent numerical solution 
of Eqs.\ (\ref{self-consistent1}-\ref{self-consistent2}) for $T_c \times n$ at $T = T_c$.
The solution of $T_c \times n$ for $V_{1,2}/t = 0.375$ is not shown in panel (b) since the 
values for $T_c$ in this case are too close to zero (coloured online).}\label{fig2}
\end{figure}

Fig.\ \ref{fig2} presents the self-consistent numerical solution of Eqs.\ (\ref{number_carriers}
-\ref{Delta_pi}) for the energy gap ratio $\Delta_2/\Delta_1$ at $T = 0\,K$ in panel (a) and 
the self-consistent numerical solution of Eqs.\ (\ref{self-consistent1}-\ref{self-consistent2}) 
for $T_c \times n$ at $T = T_c$ in panel (b). For these plots we have chosen $V_{1,1}/t = 
V_{2,2}/t = 0.50$, $\omega_c/t = 1.00$ and $V_{1,2}/t = 0.375, 0.750, 1.125, 1.500$ and $1.875$.

\section{Conclusions}\label{s5}

We have provided a theoretical description for multiband superconductivity in novel synthetic 
superlattices of superconducting layers with cubic lattice structure that will be essential for 
material design of possible room temperature superconductors \cite{bianconi93,innocenti10}.
We have investigated the two band superconductivity scenario for a two band superconducting 
metal by an extended $BCS$ mean field approach, considering the roles of the interband and 
intraband pairing. We have evaluated the critical temperature for different charge densities 
and chemical potentials, tuning the superconducting system close to relevant energies for 
electronic topological transitions. The results indicate that both interband and intraband 
effective pairing are relevant in determining the properties of the superconducting phase 
and their sensibility to the variation of external parameters ({\it i.e.}, density and chemical
potential). Interestingly, for fixed intraband pairing, and several different interband pairing, 
the critical temperature as a function of the charge density displays characteristic resonances 
(and hence amplification of the critical temperature itself) when a critical point of the band 
dispersion is crossed by the chemical potential. Therefore, the systematic analysis of the 
critical temperature and the gaps can indicate a possible root to find new synthetic 
superconducting superlattices with high values of the transition temperature. Further work 
in progress investigating the isotope effect in this complex scenario.

\begin{acknowledgments}
The authors wish to thank kindly the fol\-lowing sup\-por\-ting
Ve\-ne\-zue\-lan a\-gen\-cies: FO\-NA\-CIT (Pro\-ject S1 2002000448), 
European Community SREP project 517039 ''{\em Controlling Mesoscopic 
Phase Separation (COMEPHS)}'' and CDCH-UC (Pro\-ject 2004-014) (JJRN), and the
Bra\-zi\-lian a\-gen\-cies CNPq and FAPERGS (AAS). Numerical
calculations were performed at LANA (UFSM, Brazil) and SUPERCOMP
(UC, Venezuela). We thank M.\ D.\ Garc\'{\i}a for reading the
manuscript. Interesting discussions with C.\ I.\ Ventura,
M.\ Acquarone and R.\ Citro are fully acknowledged. Finally, one of us (JJRN) 
thanks the hospitality of ICTP Abdus Salam, Trieste, Italy as Senior 
Associate for a very productive stay (August-September 2008).
J.\ J.\ Rodr\'{\i}guez-N\'u\~nez is a member 
of the Venezuelan Scientific Program ($P.P.I.$-$IV$).
\end{acknowledgments}

\newpage

%%%%%%%%%%%%%%%% EPS FIGURES %%%%%%%%%%%%%%%%%%%%%%%%

%%%%%%%%%%%%%%%%%%%%%%%%%%%%%%%%%%%%%%%%%%%%%%%%%%%%%


\begin{thebibliography}{99}

\bibitem{SMW} H.\ Suhl, B.\ T.\ Matthias, and L.\ R.\ Walker, Phys.\ Rev.\ 
              Lett.\ {\bf 3}, 552 (1969).

\bibitem{Kondo} J.\ Kondo, Prog.\ Theor.\ Phys.\ {\bf 29}, 1 (1963).

\bibitem{Leggett} A.\ J.\ Leggett, Prog.\ Theor.\ Phys.\ {\bf 36}, 901 (1966).

\bibitem{bianconi94} A.\ Bianconi, So\-lid Sta\-te Com\-mu\-ni\-ca\-tions 89, 933 (1994), 
         ISSN 00381098, URL http://dx.doi.org/10.1016/0038-1098(94)90354-9<http://dx.doi.org/10.1016/0038-1098%2894%2990354-9.

\bibitem{bianconi05} For a review on multigap superconductivity and shape resonances:  
         A.\ Bianconi, Journal of Superconductivity and Novel Magnetism {\bf 18}, 25 (2005),
         ISSN 1557-1939, URL http://dx.doi.org/10.1007/s10948-005-0047-5.

\bibitem{bianconi93} ``{\it Process of Increasing the Critical Temperature 
        $T_c$ by Making Metal Heterostructures at the Atomic Limit}'' United States 
        Patent: US 6265019 priority date 7 Dec 1993.

\bibitem{pnictides1} Ricci, A., Joseph, B., Poccia, N., Xu, W., Chen, D., Chu, W.\ S., 
         Wu, Z. Y., Marcelli, A., Saini, N. L., and Bianconi, A., Superconductor Science and 
         Technology {\bf 23}, 052003+ (2010), ISSN 0953-2048, URL http://dx.doi.org/10.1088/0953-2048/23/5/052003.

\bibitem{agrestini} S.\ Agrestini, N.\ L.\ Saini, G.\ Bianconi, A.\ Bianconi, Journal 
         of Physics A: Mathematical and General {\bf 36}, 9133 (2003). URL http://dx.doi.org/10.1088/0305-4470/36/35/302.

\bibitem{poccian} N.\ Poccia, M.\ Fratini, Journal of Superconductivity and Novel Magnetism {\bf 22}, 299 (2009). 
         URL http://dx.doi.org/10.1007/s10948-008-0435-8.

\bibitem{pnictides2} A.\ Ricci, N.\ Poccia, B.\ Joseph, L.\ Barba, G.\ Arrighetti, G.\ Ciasca, J.\ Q.\ Yan, 
         R.\ W.\ McCallum, T.\ A.\ Lograsso, N.\ D.\ Zhigadlo, et al., Phys.\ Rev.\ B {\bf 82}, 144507+ (2010), 
         URL http://dx.doi.org/10.1103/PhysRevB.82.144507.

\bibitem{agrestini2} A.\ Bianconi, N.\ L.\ Saini, S.\ Agrestini, D.\ Di Castro, G.\ Bianconi, 
         International Journal of Modern Physics B {\bf 14}, 3342 (2000), 
         URL http://dx.doi.org/doi:10.1142/S0217979200003812.

\bibitem{agrestini3} A.\ Bianconi, S.\ Agrestini, G.\ Bianconi, D.\ Di Castro, N.\ L.\ Saini, 
         Journal of Alloys and Compounds 317-318, 537 (2001), ISSN 09258388, 
         URL http://dx.doi.org/10.1016/S0925-8388(00)01383-9.

\bibitem{kristoffel} N.\ Kristoffel, P.\ Rubin, T.\ Ord, Journal of Physics: Conference Series 108, 012034+ 
         (2008), ISSN 1742-6596, URL http://dx.doi.org/10.1088/1742-6596/108/1/012034.

\bibitem{norman10} Norman, Michael, Physics 3 (2010), ISSN 1943-2879, URL http://dx.doi.org/10.1103/Physics.3.86.

\bibitem{bianconi01} Bianconi, A., Castro, Di D., Agrestini, S., Campi, G., Saini, N.\ L., Saccone, A., Negri, 
         S.\ De, and Giovannini, M., Journal of Physics: Condensed Matter 13, 7383 (2001), ISSN 0953-8984, 
         URL http://arxiv.org/abs/cond-mat/0103211.

\bibitem{bianconi02} A.\ Bianconi , S. Agrestini, D.\ Di Castro, 
         G.\ Campi, G.\ Zangari, N.\ L.\  Saini, A.\ Saccone, S.\ De 
         Negri, M.\ Giovannini, G.\ Profeta, A.\ Continenza, G.\ Satta, S.\ 
         Massidda, A.\ Cassetta, A.\ Pifferi, M.\ Colapietro, 
         Phys.\ Rev.\ B {\bf 65}, 174515 (2002).

\bibitem{Bussmann03} A.\ Bussmann-Holder and A.\ Bianconi, Phys.\ 
         Rev.\ B {\bf 67}, 132509 (2003).
         
\bibitem{Ummarino04} G.\ A.\ Ummarino, R.\ S.\ Gonnelli, S.\ Massidda 
         and A.\ Bianconi, Physica C {\bf 407}, 121 (2004).

\bibitem{Tsuda05} S.\ Tsuda, T.\ Yokoya, T.\ Kiss, T.\ 
         Shimojima, S.\ Shin, T.\ Togashi, S.\ Watanabe, C.\ Zhang, 
         C.\ T.\ Chen, S.\ Lee, H.\ Uchiyama, S.\ Tajima, N.\ Nakai, 
         and K.\ Machida, Phys.\ Rev.\ B {\bf 72}, 064527 (2005).
         
\bibitem{Cooley} L.\ D.\ Cooley A. \ J. \ Zambano, A.\ R.\ Moodenbaugh, 
         R.\ F.\ Klie, J-C.\ Zheng, and Y. \ Zhu, Phys.\ Rev.\ Lett.\ 
         {\bf 95}, 267002 (2005).

\bibitem{Samuely} P.\ Samuely, P.\ Szab\'o, P.\ C.\ Canfield, 
         S.\ L.\ Bud'ko, Phys.\ Rev.\ Lett.\ {\bf 95}, 099701 (2005).
\bibitem{carrington} A. Carrington, J. D. Fletcher, J. R. Cooper, O. J. Taylor, L. 
         Balicas, N. D. Zhigadlo, S. M. Kazakov, J. Karpinski, J. P. H. Charmant, 
	 and J. Kortus, Phys. Rev. B {\bf 72}, 060507 (R) (2005).

\bibitem{Klie} R.\ F.\ Klie, J.\ C.\ Zhen, Y.\ Zhu, A.\ J.\ Zambano and L.\
         D.\ Cooley, Phys.\ Rev.\ B {\bf 73}, 014513 (2006); see also, T.\
         Klein, L.\ Lyard, J.\ Marcus, C.\ Marcenat, P.\ Szab\'o, Z.\
         Hol'anov\'a, P.\ Samuely, B.\ W.\ Kang, H-J.\ Kim, H-S.\
         Lee, H-K.\ Lee, and S-I. Lee, Phys.\ Rev.\ B {\bf 30}, 224528 (2006).
         
\bibitem{sologubenko1} A.\ V.\ Sologubenko, N.\ D.\ Zhigadlo, J.\ Karpinski, 
	 and H.\ R.\ Ott, Phys.\ Rev.\ B {\bf 74}, 184523 (2006).

\bibitem{castro1} D.\ Di Castro, E.\ Cappelluti, M.\ Lavagnini, A.\ 
	Sacchetti, A.\ Palenzona, M.\ Putti, and P.\ Postorino, Phys.\ Rev.\ B 
	{\bf 74}, 100505 (2006).
        
\bibitem{Putti_et_al} M.\ Putti, P.\ Brotto, M.\ Monni, Galleani d'Agliano, 
         A.\ Sanna and S.\ Massida, Europhys.\ Lett.\ \textbf{77}, 57005 (2007).

\bibitem{szabo1} P.\ Szab\'o, P.\ Samuely, Z.\ Pribulov\'a, M.\ Angst, S.\ Bud'ko, 
         P.\ C.\ Canfield, and J.\ Marcus, Phys.\ Rev.\ B {\bf 75}, 144507 (2007).

\bibitem{giubileo1} F.\ Giubileo, F.\ Bobba, A.\ Scarfato, A.\ M.\ Cucolo, 
         A.\ Kohen, D.\ Roditchev, N.\ D.\ Zhigadlo, and J.\ Karpinski,  
         Phys.\ Rev.\ B {\bf 76}, 024507 (2007).
        
\bibitem{pissas1} M.\ Pissas, D.\ Stamopoulos, N.\ Zhigadlo, and J.\ Karpinski, 
	 Phys.\ Rev.\ B {\bf 75}, 184533 (2007).

\bibitem{capua1} R.\ Di Capua, H.\ U.\ Aebersold, C.\ Ferdeghini, V.\ Fernando, 
 	 P.\ Orgiani, M.\ Putti, M.\ Salluzso, R.\ Vaglio, and 
	 X.\ X.\ Xi, Phys. Rev. B {\bf 75}, 014515 (2007).

\bibitem{hosono} Y.\ Kamihara, T.\ Watanabe, M.\ Hirano, H.\ Hosono, Journal 
         of the American Chemical Society 130, 3296 (2008), 
         URL http://dx.doi.org/10.1021/ja800073m.

\bibitem{dagotto} M.\ Daghofer, A.\ Nicholson, A.\ Moreo, E.\ Dagotto, 
         Phys.\ Rev.\ B {\bf 81}, 014511 (2010) URL http://arxiv.org/abs/0910.1573.

\bibitem{chubukov} A.\ V.\ Chubukov, D.\ V.\ Efremov, and I.\ Eremin, 
         Phys.\ Rev.\ B {\bf 78}, 134512+ (2008), URL http://dx.doi.org/10.1103/PhysRevB.78.134512.

\bibitem{fratini} Fratini, M.\ et al.\, Scale-free structural organization of oxygen 
         interstitials in la2cuo4+y. Nature 466, 841-844 (2010). URL http://dx.doi.org/10.1038/nature09260.

\bibitem{palmisano} V.\ Palmisano, L.\ Simonelli, A.\ Puri, M.\ Fratini, Y.\ Busby, P.\ Parisiades, 
         E.\ Liarokapis, M.\ Brunelli, A.\ N.\ Fitch, A.\ Bianconi, Journal of Physics: Condensed Matter 20, 
         434222+ (2008), ISSN 0953-8984, URL http://dx.doi.org/10.1088/0953-8984/20/43/434222.

\bibitem{caivano} R.\ Caivano, M.\ Fratini, N.\ Poccia, A.\ Ricci, A.\ Puri, Z.\ A.\ Ren, X.\ L.\ Dong, 
         J.\ Yang, W.\ Lu, Z.\ X.\ Zhao, et al., Superconductor Science and Technology 22, 014004+ (2009), 
         ISSN 0953-2048, URL http://dx.doi.org/10.1088/0953-2048/22/1/014004.

\bibitem{bianconi98} A.\ Bianconi, A.\ Valletta, A.\ Perali, 
         and N.\ L.\ Saini, Physica C {\bf 296}, 269 (1998). 

\bibitem{innocenti10} D.\ Innocenti, N.\ Poccia, A.\ Ricci, A.\ Valletta, S.\ Caprara, A.\ Perali, 
         and A.\ Bianconi (2010), Phys.\ Rev.\ {\bf B}, 82, 184528.

\bibitem{JJ-AAS} J.\ J.\ Rodr\'{\i}guez-N\'u\~nez and A.\ A.\ Schmidt, Phys.\ Rev.\ B {\bf 68}, 224512 (2003).
 
\bibitem{Rodriguez_2008} J.\ J.\ Rodr\'iguez - N\'u\~nez, A.\ A.\ Schmidt, A.\ Bianconi and A.\ Perali, 
         Physica C \textbf{468}, 2299 - 2304 (2008).

\bibitem{angilella} G.\ G.\ N.\ Angilella, A.\ Bianconi and R.\ Pucci, J.\ Supercond.\ {\bf 18}, 619 (2006).    

\bibitem{hexagonal} In this paper, we follow the band structure of Eq.\ (\ref{chosen_bandstructure}) that 
         is appropriate for a generic cubic lattice like for the families of iron pnictides and for quantum 
         size effects in superlattices made of layers with cubic structure. The band structure for the MgB2 
         hexagonal lattice (in 3--$D$) is given by 
\begin{eqnarray}\label{hexagonal2}
E_{1,2}(\vec{k}) &=& 
\pm V_1 \left( A(\vec{k}) + B(\vec{k}) \right)^{1/2} \nonumber \\
&+& 4~V_2~\cos(k_xa/2)~\cos (\sqrt{3}k_ya/2) \nonumber \\
&+& 2~V_3~\cos(k_zc) ~~~\\
A(\vec{k}) &\equiv& \left[ \cos(k_ya/\sqrt{3}) + 
2~\cos (k_ya/(2\sqrt{3}))~\cos (k_xa/2)
\right]^{2} \nonumber \\
B(\vec{k}) &+& \left[ 2~\sin (k_ya/(2\sqrt{3}))~
\cos (k_x a/2) - \sin (k_y a/\sqrt{3}) \right]^{2}\nonumber 
\end{eqnarray}
\noindent where $V_{1,2}$ is the hopping integral between atoms placed on 
the hexagonal lattice (second type of atoms placed inside the primitive cell). $V_3$ is the 
effective hopping integral along the $c$--axis (or $z$--axis). The hexagonal lattice 
(Eq.\ (\ref{hexagonal2})) is  left for a future calculation.

\end{thebibliography}
\end{document}